\documentclass[conference]{IEEEtran}
\IEEEoverridecommandlockouts
\usepackage{cite}
\usepackage{amsmath,amssymb,amsfonts}
\usepackage{algorithmic}
\usepackage{graphicx}
\usepackage{textcomp}
\usepackage{xcolor}

\usepackage{booktabs} 

\def\BibTeX{{\rm B\kern-.05em{\sc i\kern-.025em b}\kern-.08em
    T\kern-.1667em\lower.7ex\hbox{E}\kern-.125emX}}
\begin{document}

\title{Quantum Annealing and Graph Neural Networks for Solving TSP with QUBO
}

\author{\IEEEauthorblockN{1\textsuperscript{st} Haoqi He}
\IEEEauthorblockA{\textit{Sun Yat-sen University} \\
\textit{Cybersecurity school}\\
Shenzhen, China \\
0009-0007-8646-8593}

}
\maketitle

\begin{abstract}
This paper explores the application of Quadratic Unconstrained Binary Optimization (QUBO) models in solving the Travelling Salesman Problem (TSP) through Quantum Annealing algorithms and Graph Neural Networks. Quantum Annealing (QA), a quantum-inspired optimization method that exploits quantum tunneling to escape local minima, is used to solve QUBO formulations of TSP instances on Coherent Ising Machines (CIMs). The paper also presents a novel approach where QUBO is employed as a loss function within a GNN architecture tailored for solving TSP efficiently. By leveraging GNN's capability to learn graph representations, this method finds approximate solutions to TSP with improved computational time compared to traditional exact solvers.

The paper details how to construct a QUBO model for TSP by encoding city visits into binary variables and formulating constraints that guarantee valid tours. It further discusses the implementation of QUBO-based Quantum Annealing algorithm for TSP (QQA-TSP) and its feasibility demonstration using quantum simulation platforms. 

In addition, it introduces a Graph Neural Network solution for TSP (QGNN-TSP), which learns the underlying structure of the problem and produces competitive solutions via gradient descent over a QUBO-based loss function.

The experimental results compare the performance of QQA-TSP against state-of-the-art classical solvers such as dynamic programming, Concorde, and Gurobi, while also presenting empirical outcomes from training and evaluating QGNN-TSP on various TSP datasets. The study highlights the promise of combining deep learning techniques with quantum-inspired optimization methods for solving NP-hard problems like TSP, suggesting future directions for enhancing GNN architectures and applying QUBO frameworks to more complex combinatorial optimization tasks.
\end{abstract}


\section{Introduction}
QA simulates the "quantum annealing" phenomenon from quantum mechanics. During its evolution process, the system's Hamiltonian becomes minimized, indicating that the system transitions gradually from an initial state to its ground state within the context of quantum states\cite{b1}. Decoding this ground state reveals the global minimum energy or optimal solution for the given problem.

The QA algorithm harnesses the superposition and entanglement properties of Qubits, enabling the system to perform parallel searches across all possible solutions. 

By leveraging quantum tunneling effects, it can potentially overcome local minima barriers, thus expediting the discovery of the global optimum solution\cite{b2,b11}.

Thanks to advancements in CIMs for quantum annealing and the continuous increase in the number of Qubits, utilizing quantum annealing for accelerated computations is steadily becoming a practical reality\cite{b3,b14}.

ML and DL offer powerful solving capabilities across various domains; however, these algorithms require training before implementation and deployment – a process involving tuning model parameters to extract meaningful information from data\cite{b5}. GNNs demonstrate structural adaptability compared to traditional neural networks when handling complex structured data. Essentially, GNNs can directly model graph structures and possess the potential to better capture relational information within graphs\cite{b8}.

Quantum Annealing Algorithms and Graph Neural Network models represent emerging approaches for tackling combinatorial optimization problems such as the TSP\cite{b9,b16,b17}.

The TSP is a classic example in combinatorial optimization where, given a weighted connected graph \(G(V,E)\), the task is to traverse all vertices exactly once and return to the starting point with the shortest total distance traveled. This scenario represents a situation where a salesman departs from one city, visits a series of other cities, and finally returns to the originating city, with the distances (or costs, times, etc.) between each city being known. The challenge lies in finding the sequence of visits that minimizes the overall journey distance\cite{b13}.

\begin{figure}[htbp]
  \centering
  \includegraphics[width=0.5\textwidth]{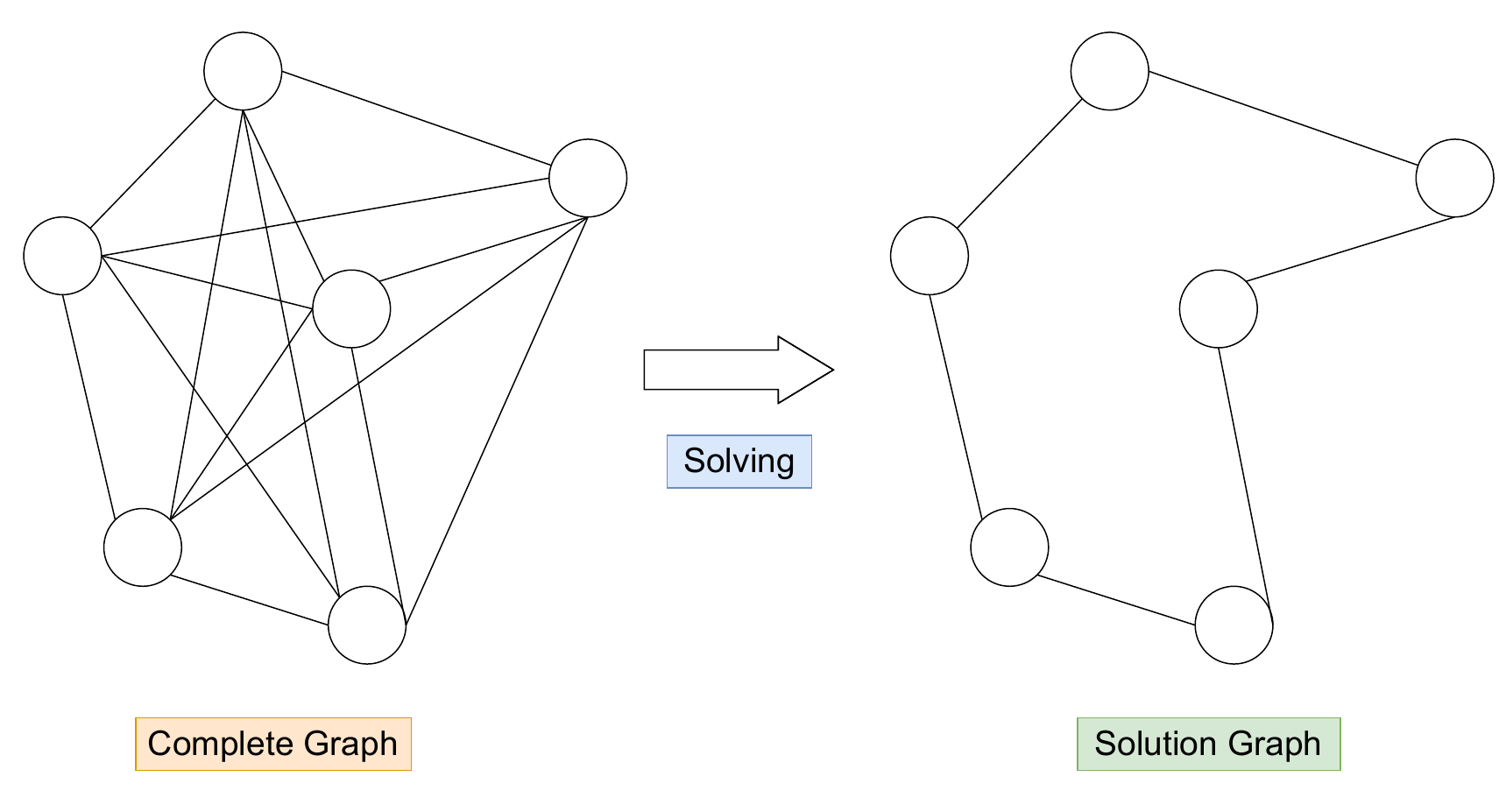}
  \caption{TSP example}
  \label{fig:example}
\end{figure}

A broad range of applications, from logistics to electronics manufacturing, can leverage TSP solutions for optimizing resource-constrained visitation or path planning\cite{b18,b26,b27}. In logistics and distribution, businesses seek the most efficient delivery routes that minimize total travel distance or time between warehouses and multiple client destinations. 

Meanwhile, in electronics design, engineers face challenges akin to TSP when they aim to reduce signal transmission delays in printed circuit boards.

The TSP stands out as one of the most intensively studied NP-hard problems in the field of combinatorial optimization. NP-hard problems exhibit "Optimality Complexity," implying that their optimal solutions may demand exponential or even worse computational complexity. With \(n\) cities in the TSP, the total number of possible solutions amounts to \(n!\) combinations, rendering exact solutions impractical for moderately large \(n\).\cite{b13} Despite the inherent difficulty, the broad applicability of TSP has made obtaining approximate solutions a pressing need. 

In practical applications, we can utilize QA and GNNs to derive approximate yet accurate solutions within an acceptable timeframe\cite{b4,b6,b7}.

QUBO modeling is a mathematical optimization technique that proves particularly useful in the realm of quantum computing because it can be directly mapped onto the Ising Model – a physical system that can be simulated by CIMs.

By mapping these problems onto the Ising model, they can be solved using quantum annealing or other methods that exploit quantum properties\cite{b1,b3}. In the QUBO model, the objective is to find the optimal configuration of a set of binary variables (where each variable can only take values 0 or 1), thereby minimizing a specific quadratic polynomial function.
\begin{figure}[htbp]
  \centering
  \includegraphics[width=0.51\textwidth]{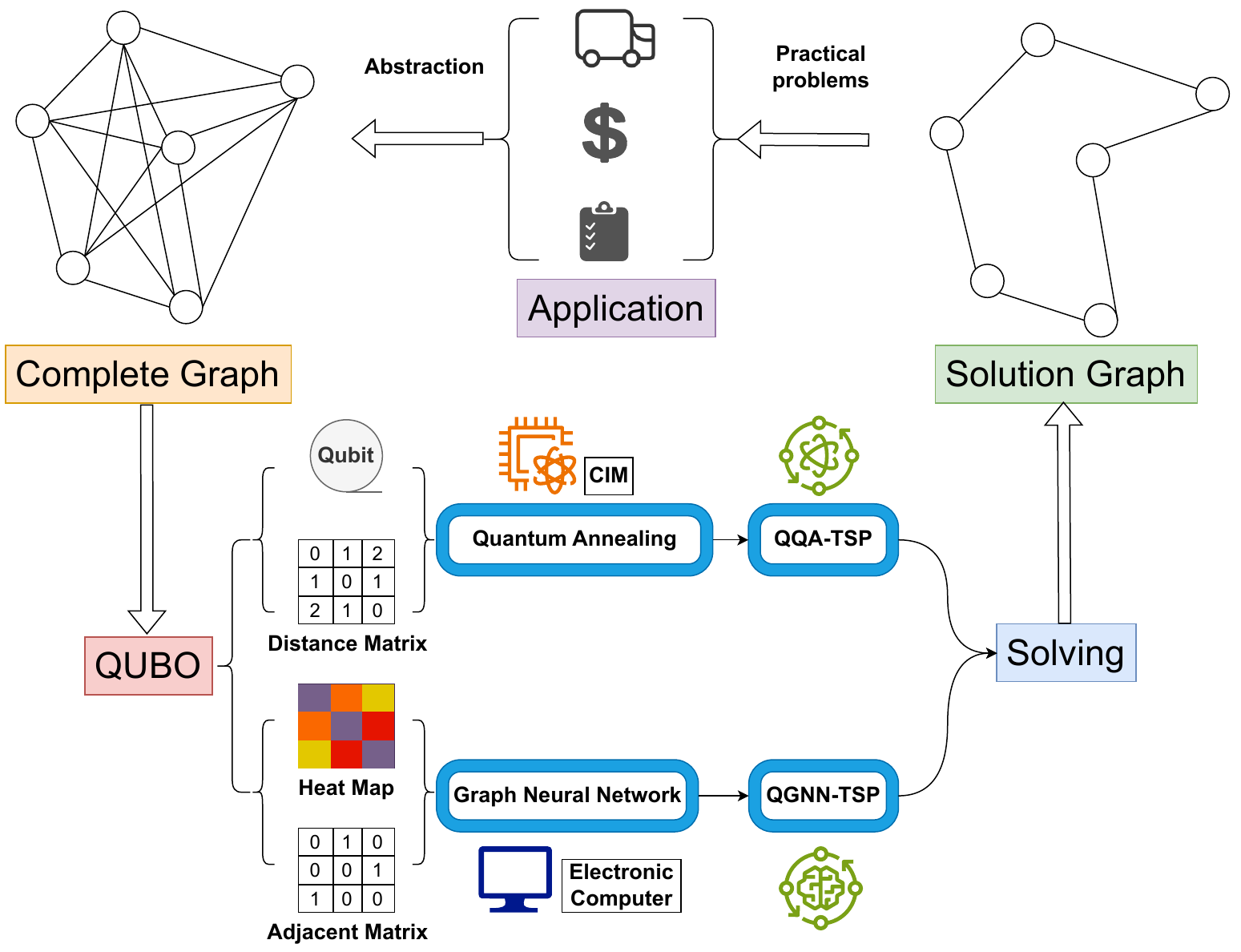}
  \caption{Models we have promoted}
  \label{fig:example}
\end{figure}

The primary contributions of this paper include: describing the application of QUBO in modeling TSP; proposing QQA-TSP from the perspective of quantum annealing and proving its feasibility through simulations; and introducing QGNN-TSP from the standpoint of neural networks, showcasing its superiority over traditional exact algorithms in terms of time efficiency.

\section{QUBO in TSP}

\subsection{QUBO modeling}
By constructing the matrix $\mathbf{Q}$, we can transform various practical problems into QUBO form.
\begin{equation}
    \text {$ \mathbf{QUBO} =  x^T\mathbf{Q}x \ \ \ \ \ x_{i}  \in \{0 \ or \ 1\}^n $}
\end{equation}

In the context of TSP, a QUBO model can be formally represented by the following mathematical expression:

$$
    \mathbf{H} = \sum_{i=0}^{n-1} Q_{i,i} x_i^2 + \sum_{i=0}^{n-2}\sum_{j=i+1}^{n-1} Q_{i,j} x_i x_j 
$$

Alternatively, this can be expressed as:
\begin{equation}
    \mathbf{H} = \sum_{i=0}^{n-1}\sum_{j=0}^{n-1} Q_{i,j} x_{i,j} 
\end{equation} 
where \( \mathbf{x} \) is an n-dimensional binary vector with components \( x_i \in \{0, 1\} \) for \( i = 0, 1, \ldots, n-1 \).

The matrix \( \mathbf{Q} \) is a symmetric \( n \times n \) matrix, and its elements \( Q_{ij} \) denote the interaction coefficients between the binary variables \( x_i \) and \( x_j \)\cite{b10}.
To transform TSP into a QUBO model, adjacency and distance matrices play a crucial role. Given a weighted graph \( G(\mathbf{V},\mathbf{E}) \), where \( \mathbf{V} \) represents the set of all city nodes, and \( \mathbf{E} \) denotes the set of edges connecting cities, each edge \( (u,v) \in \mathbf{E} \) has a weight \( D_{u,v} \) signifying the distance from city \( u \) to city \( v \).

The Encoding Scheme used to solve the TSP is as follows. For every city \( u \) and each possible visitation order \( j \), we define a corresponding binary variable \( x_{u,j} \). 

If city \( u \) is visited at position \( j \) in the tour, then \( x_{u,j} = 1 \); otherwise, \( x_{u,j} = 0 \). This way, any feasible tour can be encoded as a specific combination of these binary variables, effectively constructing an extended adjacency matrix\cite{b12,b18,b19}.

To ensure that the model complies with TSP requirements, two types of constraints are imposed:

   Each city appears only once in the path; thus, for every city \( u \in \mathbf{V} \), the sum of its visitation sequence should equal 1 as the first uniqueness constraint:
   \begin{equation}
        \sum_{j=0}^{N-1}x_{u,j}=1, \quad \forall u \in \mathbf{V}
   \end{equation}

   Similarly, each position in the sequence corresponds to exactly one visited city as the second uniqueness constraint as:
   \begin{equation}
       \sum_{u=0}^{N-1}x_{u,j}=1, \quad \forall j \in \{0, ..., N-1\}
   \end{equation}

   For non-adjacent cities \( u \) and \( v \), they cannot appear consecutively in the path, which can be mathematically written as:
   \begin{equation}
     Con = \sum_{(u,v)\notin \mathbf{E}}\left(\sum_{j=0}^{N-2}x_{u,j}x_{v,j+1} + x_{u,N-1}x_{v,0}\right)   
   \end{equation}
  
The above constraints are converted into quadratic terms and incorporated into the QUBO objective function \( \mathbf{H} \), aiming to minimize the total path length. The constraint function is specified as:

\begin{equation}
\mathbf{F} = \sum_{u=0}^{N-1} \left(\sum_{j=0}^{N-1}x_{u,j}-1\right)^2 + \sum_{j=0}^{N-1} \left(\sum_{u=0}^{N-1}x_{u,j}-1\right)^2 + Con
\end{equation}

The cost function can be expressed as the sum of products involving the distance matrix and the binary variables representing city connections:
\begin{equation}
   \mathbf{C} = \sum_{(u,v)\in \mathbf{E}}D_{u,v}\left(\sum_{j=0}^{N-2}x_{u,j}x_{v,j+1} + x_{u,N-1}x_{v,0}\right)
\end{equation}

Ultimately, the goal is to satisfy the constraints while minimizing the cost. As both the constraints and the cost are inherently positive, the Hamiltonian to compute becomes:
\begin{equation}
    \mathbf{H} = A \cdot \mathbf{F} + \mathbf{C}
\end{equation}

Here, the coefficient \( A \) serves to balance the penalty incurred when violating constraints against the actual cost of the path length. By adjusting this coefficient, a compromise can be reached between minimizing the probability of constraint violation and discovering shorter paths.

\subsection{QUBO based Quantum Annealing algorithm Solving TSP}
In the preceding text, a mathematical analysis of the Hamiltonian has been conducted. Let us now elucidate how QQA-TSP is implemented on a CIM to expedite the solution process.

The CIM's implementation of quantum annealing demonstrates significant advantages. It can emulate the collective dynamics of quantum systems, particularly when addressing large-scale optimization problems. The CIM enables rapid parallel search and under certain conditions, exhibits superior performance compared to classical computers by swiftly converging to an approximate optimal solution\cite{b2,b3}. Notably, the Ising model can be directly mapped onto the CIM without requiring additional encoding procedures.

The Ising model, mathematically equivalent to the QUBO model and convertible between them, is typically used to describe spin interactions in magnetic systems. The Hamiltonian for the Ising Model takes the form:

\begin{equation}
    H_{\text{Ising}} = -\sum_{i=1}^{N} h_i s_i - \sum_{i<j}^{N} J_{i,j} s_i s_j 
\end{equation}

Here, \( s_i \) represents the \( i^{\text{th}} \) spin variable, taking values of either \( +1 \) or \( -1 \),  \( h_i \) denotes the local magnetic field strength applied to the \( i^{\text{th}} \) spin, \( J_{i,j} \) signifies the coupling strength between the \( i^{\text{th}} \) and \( j^{\text{th}} \) spins, \( N \) stands for the total number of spins.

The transformation from QUBO to Ising models can be achieved through the following mapping:

\begin{equation}
    s_i = (2x_i - 1) 
\end{equation}

Substituting this into the Ising Hamiltonian yields the corresponding relationship:

\begin{equation}
    H_{\text{Ising}} = -\sum_{i=1}^{N} h'_i (2x_i - 1) - \sum_{i<j}^{N} J'_{ij} (2x_i - 1)(2x_j - 1) 
\end{equation}

By adjusting coefficient relations such that \( h'_i \) and \( J'_{ij} \) correspond to the original \( h_i \) and \( J_{ij} \), respectively, we complete the conversion from the QUBO model to the Ising model.

At the heart of quantum annealing lies the construction of a time-dependent Hamiltonian \( H(t) \)\cite{b15}, which defines the system's total energy in a quantum setting and evolves from an easily manageable initial state to one that incorporates information about the optimization problem at hand. Typically, this Hamiltonian decomposes into two components:

Driving Hamiltonian (also known as Control or Kinetic Hamiltonian) \( H_{kin}(t) \) describes quantum interactions and coherence among qubits, enabling the system to exist in superposition states and tunnel through classical barriers.
Problem Hamiltonian(Target or Potential Hamiltonian) \( H_{pot}(t) \) corresponds to the objective function to be optimized and is often represented in QUBO forms. 

The entire quantum annealing process can be described by the Schrödinger equation\cite{b20}:

\begin{equation}
  i\hbar \frac{\partial}{\partial t}|\psi(t)\rangle = H(t)|\psi(t)\rangle  
\end{equation}

where \( |\psi(t)\rangle \) represents the wavefunction evolving over time, and \( H(t) \) initially predominantly reflects \( H_{kin}(t) \), gradually transitioning to \( H_{pot}(t) \) during the annealing process\cite{b12,b15}.
Upon reaching the endpoint, the system cools down to its ground state, where the probability distribution of the qubits becomes concentrated at the lowest energy state of \( H_{pot}(t) \), corresponding to the global or near-global minimum of the target function.

The central role is played by the Schrodinger equation, which describes how quantum systems evolve over time. The Hamiltonian consists of two components: the kinetic energy term \(H_{kin} \) and the potential energy term \(H_{pot} \), corresponding to an initial easily decoded state and the final objective optimization problem form, respectively.

For the TSP problem, we can construct a Hamiltonian of the following form:

\begin{equation}
H(t) = A(t)H_{kin} + B(t)H_{pot},
\end{equation}

Where \(A(t) \) and \(B(t) \) are time-dependent coefficients, initially \(A(0) \gg B(0) \), so that the system is mainly controlled by the kinetic energy term, as the quantum annealing process proceeds, \(A(t) \) slowly decreases and \(B(t) \) gradually increases. Up to \(t=t_f \) \(A(t_f) \approx 0 \) and \(B(t_f) \approx 1 \), the system is almost completely influenced by the potential energy term corresponding to the objective optimization problem.

The Schrodinger equation in this case is expressed as follows.

\begin{equation}
i\hbar \frac{\partial}{\partial t}|\psi(t)\rangle = \left[A(t)H_{kin} + B(t)H_{pot}\right]|\psi(t)\rangle,
\end{equation}

Along with the time evolution, the wave function \( | \psi (t) \ rangle \) will tend to be at the lowest energy state of the whole system, especially after the annealing. At this point, the state of the system is concentrated in the ground state of \(H_{pot}(t_f) \), which corresponds to the global or near-global optimal solution of the original optimization problem.

In summary, in the process of quantum annealing, the Schrodinger equation is used to drive the quantum state to lower energy distribution, so as to effectively find the optimal solution or near-optimal solution of complex combinatorial optimization problems such as TSP.

\begin{figure}[htbp]
  \centering
  \includegraphics[width=0.5\textwidth]{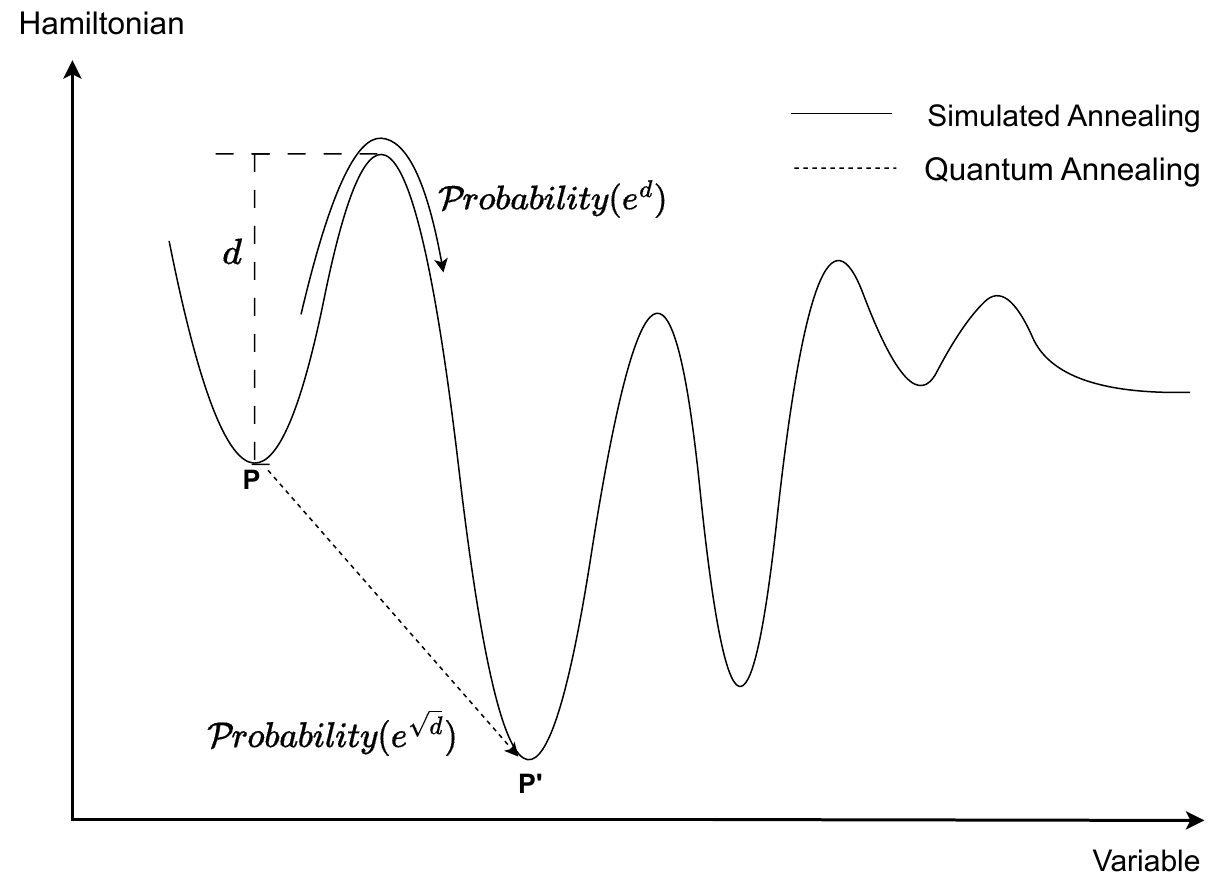}
  \caption{Simulated annealing and Quantum annealing are compared under escape suboptimal solutions}
  \label{fig:example}
\end{figure}

Once the minimum Hamiltonian is obtained, it can be decoded to reveal the optimal solution.

Regarding complexity analysis, the QUBO model we have proposed necessitates a qubit space complexity of \( \mathcal{O}(N^2) \).

The reason why QA has \( \mathcal{O}(N^2) \) Qubit space complexity when solving TSP is that it needs to encode all possible states of the problem. 

For a problem like TSP, each city node can be viewed as a binary variable whose value indicates whether it is included in the final path or not. 

When there are n cities, we need n Qubits to represent the states of each of the n cities.

However, the key is not only the number of cities, but also the connection between them.

To fully express all possible combinations of paths, that is, whether a valid connection is formed between each pair of cities, it is usually necessary to consider the existence of edges between each pair of cities. Thus, in the process of constructing a Hamiltonian loop or finding an optimal path, each pair of cities requires a Qubit to represent the connected state between them, leading to an additional  \(n (n - 1) \) possible combinations of edges.

Since complete graphs are usually used in the practical solution process, the total number of qubits needed to be stored is the sum of the connection states between each pair of cities, so that all possible path permutations can be fully described. The QUBO model constructed in this way maps the whole problem onto the quantum system, so that QA can search and find the global minimum energy state, which is the optimal solution or near-optimal solution of the corresponding problem.

The time complexity for obtaining exact solutions via quantum annealing is theoretically characterized by an exponential growth rate, given as \( \mathcal{O}(e^{\sqrt{d}}) \) , where \( d \) typically represents some problem-dependent parameter\cite{b2,b4}. However, it is worth noting that the time complexity for attaining approximate solutions using statistical techniques could potentially diminish to polynomial levels, which can be expressed mathematically as \( \mathcal{O}(p(d)) \), where \( p(d) \) is a polynomial function of \( d \).

This reduction in time complexity when approximating solutions with quantum annealing stands in contrast to thermal annealing's theoretical time complexity, which is also exponential but generally steeper, represented as \( \mathcal{O}(e^{d}) \)\cite{b21}. 

This comparison highlights a potential advantage of quantum annealing over its classical counterpart in terms of efficiency for certain problems, especially when dealing with large-scale optimization instances and seeking near-optimal solutions.

\subsection{QUBO as loss function in GNN Solving TSP}
We present a GNN architecture that effectively captures local and global structural information among nodes, with node features iteratively updated through message passing. This design enables each node's state to encapsulate both its inherent properties and the collective knowledge from neighboring nodes\cite{b5,b8,b30}. In a novel integration, we embed the QUBO framework within the GNN setting for TSP resolution. By devising a loss function grounded in the QUBO formulation of TSP, our methodology ingeniously leverages GNNs' strong representation learning powers to expedite solving combinatorial optimization problems.
%
%
\begin{figure}[htbp]
  \centering
  \includegraphics[width=0.5\textwidth]{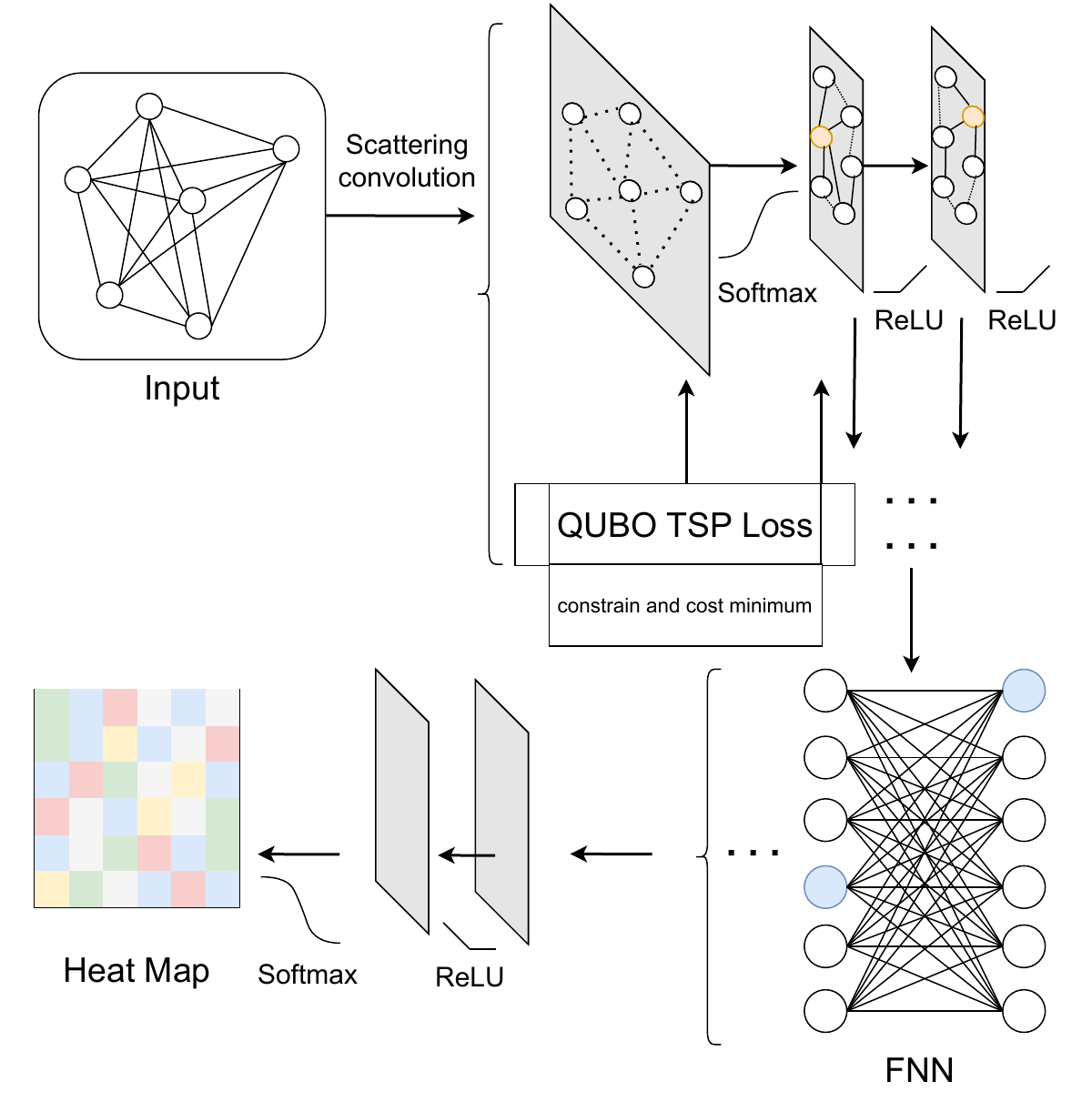}
  \caption{QUBO as loss function in GNN}
  \label{fig:example}
\end{figure}

Given a TSP instance, let \( D_{ij} \) denote the Euclidean distance between cities \( i \) and \( j \), with \( D \in \mathbb{R}^{n \times n} \) being the distance matrix. We initially construct an adjacency matrix \( W \in \mathbb{R}^{n \times n} \) with \( W_{i,j} = e^{-D_{ij}/\tau} \), embodying a scattered attention-based representation\cite{b9,b16,b24}. The node feature matrix is \( F \in \mathbb{R}^{n \times 2} \) with \( F_i = (x_i, y_i) \), where \( \tau \) represents a temperature parameter.

The node feature matrix \( F \) and weight matrix \( W \) are fed into the GNN to generate a transition matrix \( T \in \mathbb{R}^{n \times n} \). The output matrix \( S \in \mathbb{R}^{n \times n} \) from the GNN undergoes column-wise softmax activation such that \( T_{i,j} = \frac{e^{S_{i,j}}}{\sum_{k=1}^n e^{S_{k,j}}} \)\cite{b25}, ensuring row and column sums equal to 1 and non-negativity of elements\cite{b16}, akin to constraints in the problem domain.

\begin{figure}[htbp]
  \centering
  \includegraphics[width=0.5\textwidth]{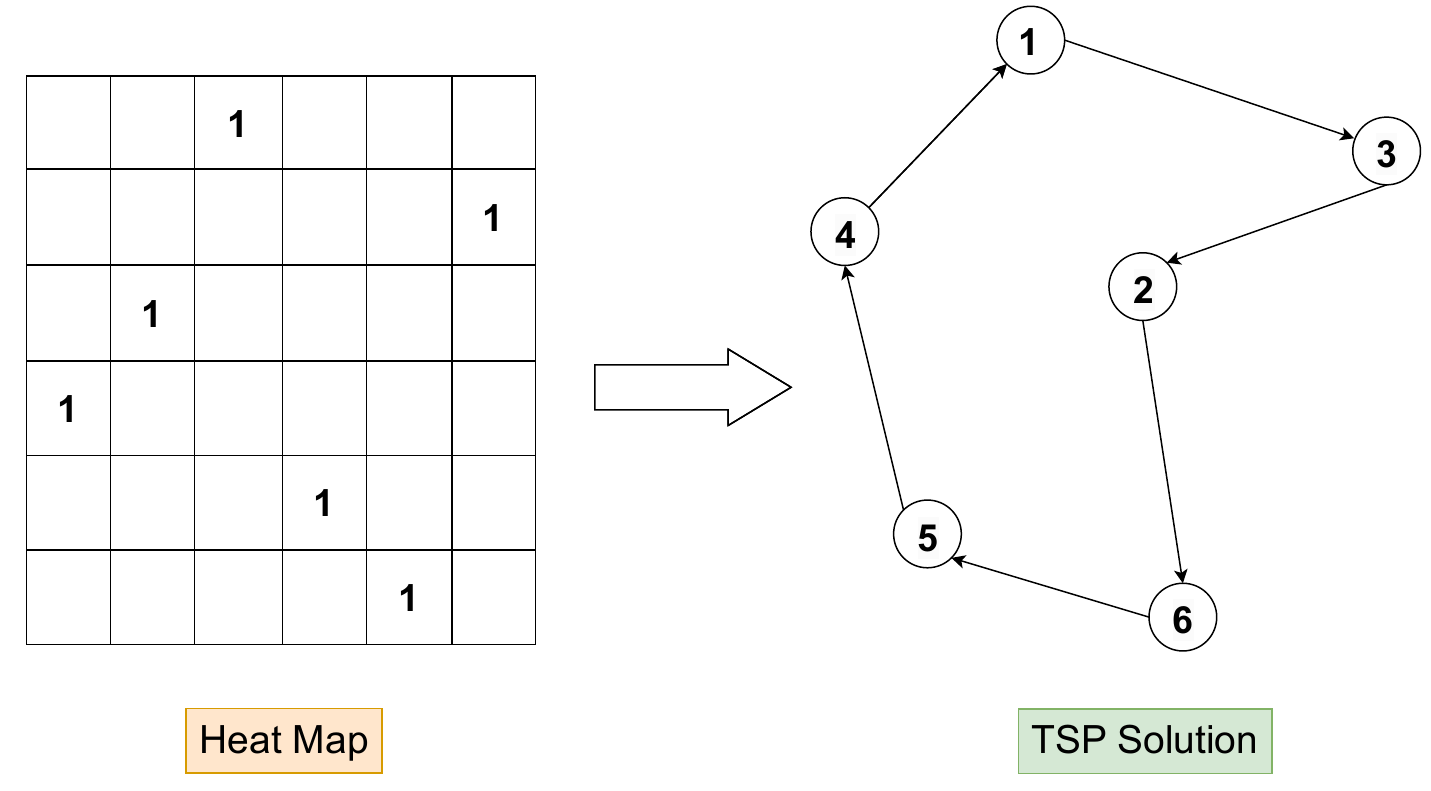}
  \caption{Heat Map and Solution}
  \label{fig:example}
\end{figure}

Subsequently, utilizing a Sylvester transformation matrix technique—although specific details are not provided here—we decode \( T_{i,j} \) into a heatmap matrix \( H \) that encodes potential TSP paths probabilistically\cite{b28}; higher probabilities correspond to nodes more likely to be included in the optimal path\cite{b17}.

Crucially, the essence of our proposition lies in employing a QUBO-based loss function during the training process as a metric measuring the discrepancy between the model's predicted tour and the optimal TSP solution. In this setup, each variable in the QUBO formulation corresponds to a decision variable in TSP—namely, whether city \( i \) is part of the tour—and the coefficients reflect the distance costs between cities. By setting the loss function to minimize the energy function associated with the TSP instance, the GNN's learning objective becomes finding low-energy states that adhere to the TSP constraints, thereby effectively encoding the problem.
$$
\mathbf{Loss_{TSP}} = \sum_{u=0}^{N-1} \left(\sum_{j=0}^{N-1}x_{u,j}-1\right)^2 + \sum_{j=0}^{N-1} \left(\sum_{u=0}^{N-1}x_{u,j}-1\right)^2 +
$$
\begin{equation}
A \sum_{(u,v)\notin \mathbf{E}}\left(\sum_{j=0}^{N-2}x_{u,j}x_{v,j+1} + x_{u,N-1}x_{v,0}\right)
\end{equation}
During practical implementation, the GNN first generates a probability distribution over whether each edge is included in a potential tour. This distribution is then mapped onto the QUBO decision variable space to calculate the corresponding QUBO loss value. Through backpropagation and gradient descent-based optimization algorithms, the GNN parameters are updated to learn latent relationship features among nodes, progressively converging towards better solutions.
%
%

Finally, using the obtained heatmap \( Heat \), a guided local search is performed to derive the final solution.

An iterative expansion method constructs and optimizes potential tours by starting with a randomly selected vertex \( u_1 \) and iteratively adding vertices \( v_i \) to form a closed loop \( v_{k+1} = u_1 \). At each expansion step, the next vertex \( u_{i+1} \) is chosen based on the following principles:
If selecting \( u_{i+1} \) results in a valid and improved TSP tour, accept this update as a candidate solution.
When the number of expansion steps \( i \) reaches a hyperparameter \( K \), indicating the maximum allowed edge removals, terminate the current expansion and restart. Otherwise, randomly select \( u_{i+1} \) from the top \( M \) candidate cities connected to the current city \( u \) based on adjusted probabilities \( L_{u,v} \), giving preference to those with higher likelihood or closer proximity according to the heatmap\cite{b22}.
The algorithm uses a cost function \( C \) as an evaluation metric, consistently seeking shorter path lengths for new solutions. 

For each search node, up to \( T \) extension attempts are made to improve solution quality. 

If no improvement is found after \( T \) tries, a new random initial solution is generated, and the best-first local search is repeated until a stopping criterion is met\cite{b23}.

\section{EXPERIMENT}

\subsection{Data set and software description }
	
In the experimental phase of this study, a series of datasets were utilized with varying sample sizes: n = 20, 50 and n = 100. 

Limited by the large memory footprint of simulating Qubit, and this paper only proves the feasibility of QQA, only n = 100 experiments are done here.

The architecture and training of the neural network models were implemented using the Pytorch framework. Furthermore, to emulate quantum annealing processes, we made use of MATLAB's dedicated quantum computing toolbox. In the context of some traditional algorithms under comparison, their corresponding results were sourced directly from existing literature without necessitating reimplementation or recalculation.

We use the same test dataset in \(Fuet al., 2021\)\cite{b31}.

\subsection{Hamiltonian iteration and Heat map}
In this section, we give a visualization of the Hamiltonian evolution of QQA and the heat map of QGNN. Hamiltonian reflects that QA gradually finds a reasonable solution in many solution Spaces, and obtaining heat map is an important step of QGNN to solve TSP.

By solving the simulated QA, we are able to obtain the evolution of qubits and Hamiltonians in QA. After about 600 iterations in the figure, the qubits diverge and tend to converge, and the Hamiltonian converges decreasingly to the interval, indicating that QA has successfully found a feasible solution.
\begin{figure}[htbp]
  \centering
  \includegraphics[width=0.5\textwidth]{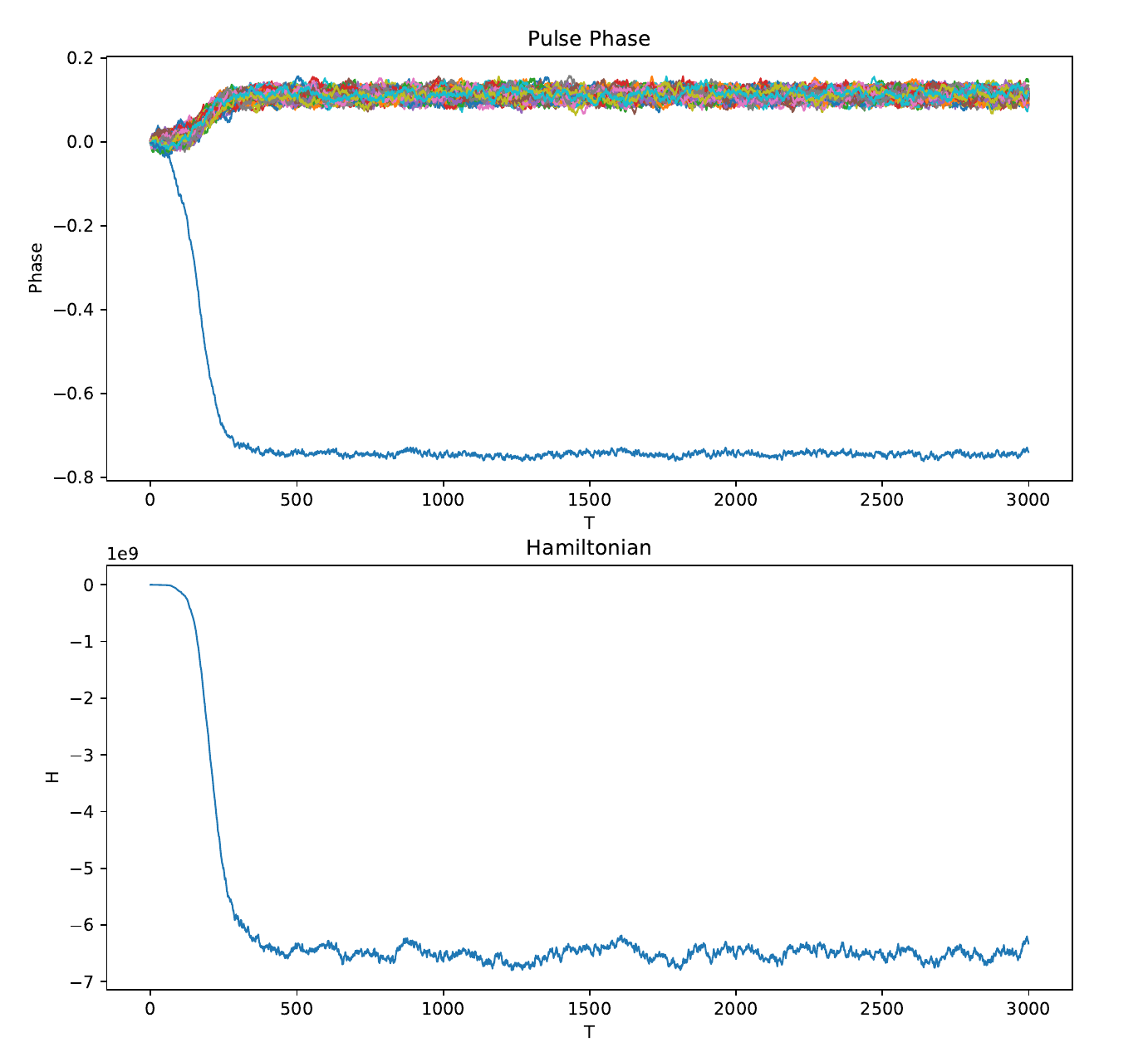}
  \caption{Iteration of QUBO loss function in QGNN}
  \label{fig:example}
\end{figure}
The initial heatmap is transformed by distance matrix, adjacency matrix and normalized to assign a uniform probability distribution on all edges.

\begin{figure}[htbp]
  \centering
  \includegraphics[width=0.5\textwidth]{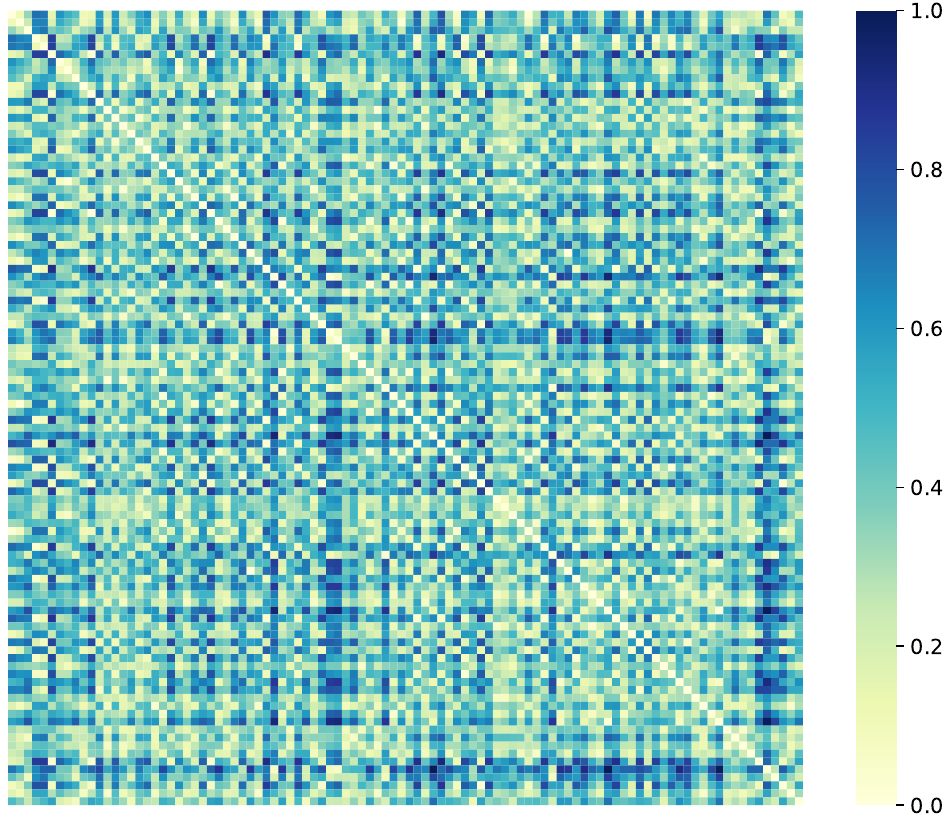}
  \caption{The initial Heat Map}
  \label{fig:example}
\end{figure}
To optimize the heatmap to better match the optimal solution, we introduce the QUBO loss function. 

In this context, the QUBO loss may be defined as a measure of the gap between the current predicted solution and the ideal TSP path length. With each epoch of training, the model updates the weights by gradient descent algorithm, which gradually decreases the value of the loss function, meaning that the predicted heatmap is closer to the state that can produce the shortest tour path.

With each epoch of training, the model updates the weights by gradient descent algorithm, which gradually decreases the value of the loss function, meaning that the predicted heatmap is closer to the state that can produce the shortest tour path. 

If the heatmap predicted by the model fails to correctly emphasize those edges that make up the shorter TSP path, the loss function will be higher.

\begin{figure}[htbp]
  \centering
  \includegraphics[width=0.5\textwidth]{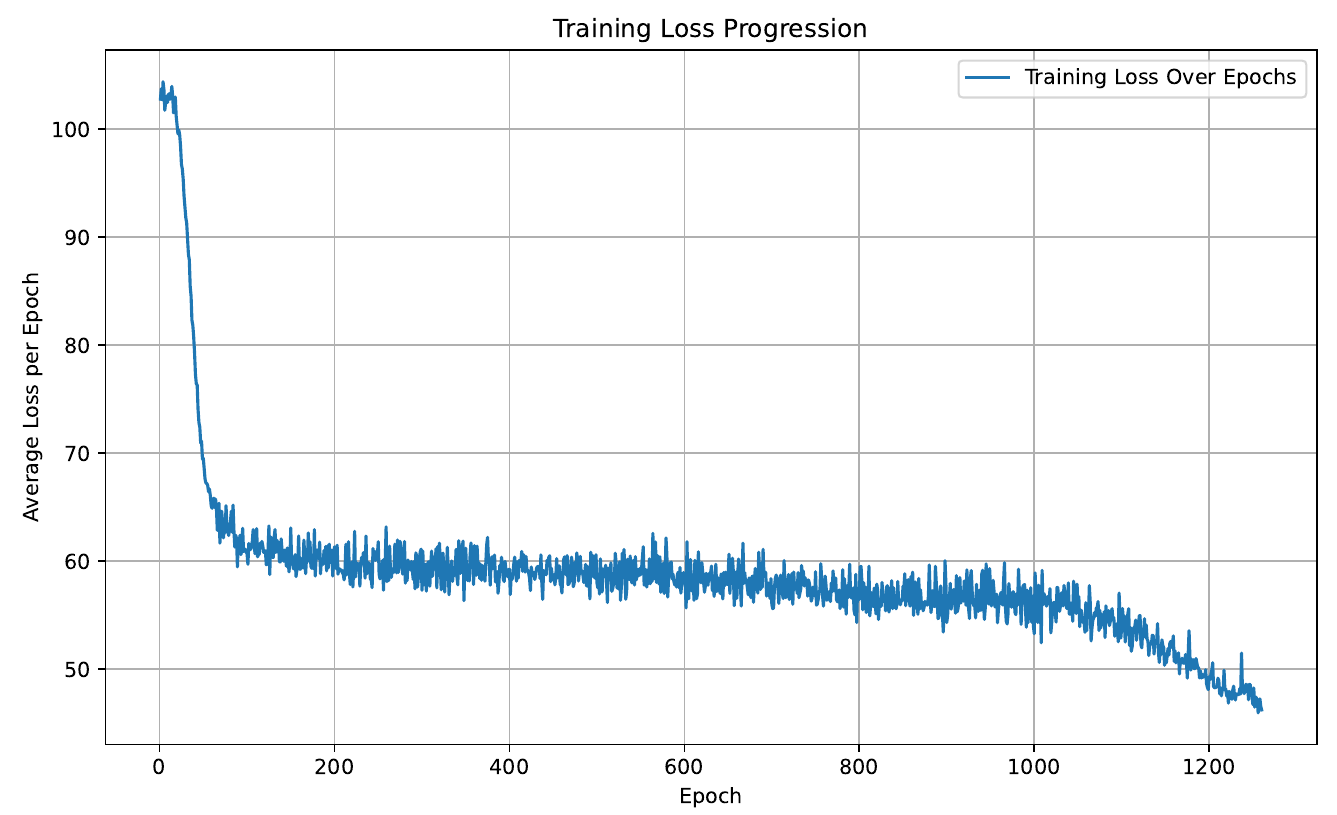}
  \caption{Evolution of qubits and Hamiltonians in QA}
  \label{fig:example}
\end{figure}
For example, during each backpropagation, the loss function will be higher if the heatmap predicted by the model fails to correctly emphasize those edges that make up the shorter TSP path. Through training, QGNN continuously adjusts its parameters to reduce this loss, thereby improving the quality of the predicted heatmap.

Through training, QGNN continuously adjusts its parameters to reduce this loss, thereby improving the quality of the predicted heatmap. When the loss function gradually decreases with epoch, it means that the model estimates the optimal solution more and more accurately.

\begin{figure}[htbp]
  \centering
  \includegraphics[width=0.48\textwidth]{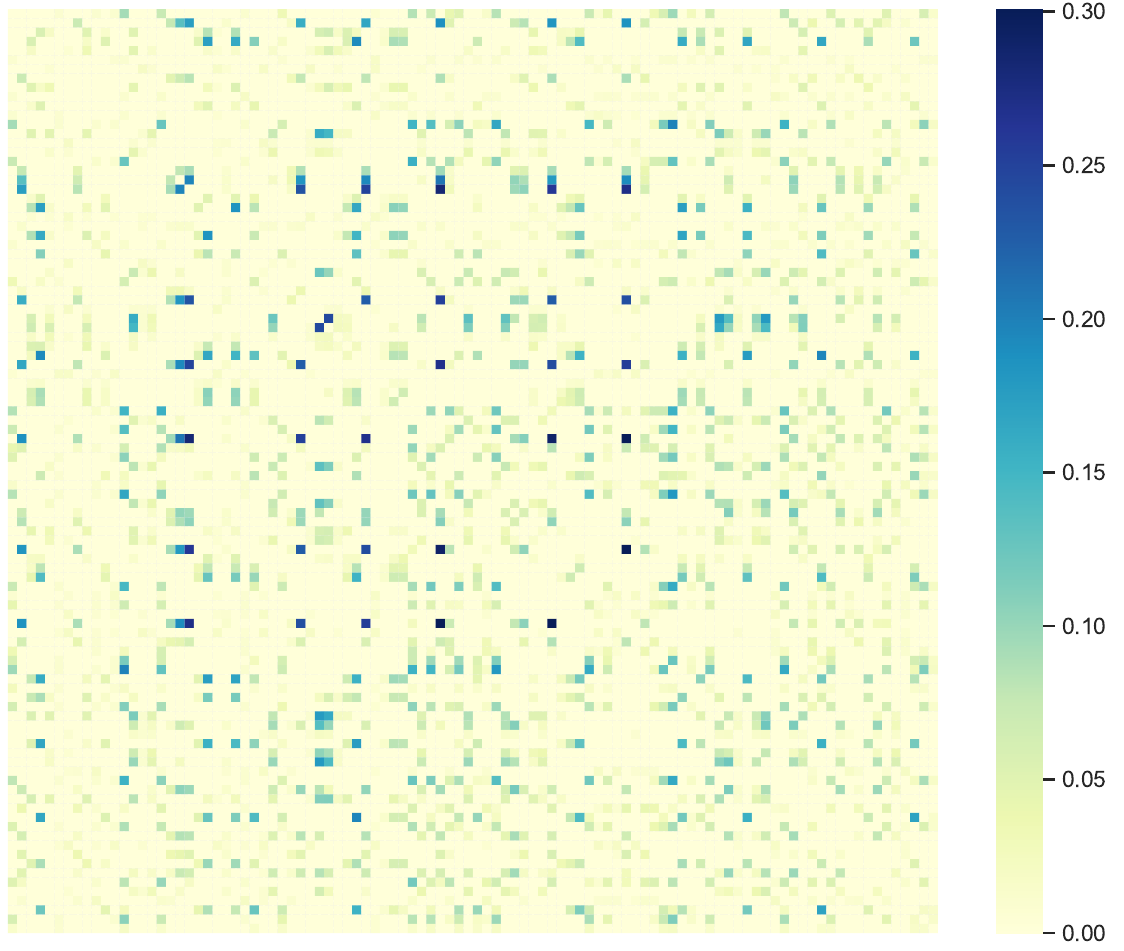}
  \caption{Heat Map After training}
  \label{fig:example}
\end{figure}
When solving TSP, an effective heatmap should be able to assign a high probability weight to the connections between each pair of cities, especially those edges that eventually constitute an optimal path or a high-quality solution. The trained heat maps obviously change from a relatively uniform probability distribution to different ones. 

In the trained heatmap, the connections between city pairs that are closer to the actual optimal path should have higher heat values. 

These high-heat edges are more likely to be included in the final generated tour path. 

This strongly verifies the superiority and feasibility of the adopted graph neural network method based on QUBO loss function in solving TSP.


\subsection{Experimental results for QQA and QGNN}
The experimental evaluation of the QQA-TSP was conducted on a range of TSP instances, transformed into QUBO formulations. The results showed that QQA-TSP achieved competitive solution quality compared to classical solvers such as dynamic programming and Concorde, particularly in small to medium-sized problem instances. For larger problems, while the absolute optimality gap increased due to hardware limitations, QQA-TSP still demonstrated a promising trend in finding near-optimal solutions within reasonable timeframes.

In contrast, the Graph Neural Network employing QUBO loss function, or QGNN-TSP was trained and tested on various TSP datasets. This approach integrated the QUBO formulation within the learning process, allowing the GNN to generate approximate solutions through gradient descent optimization.

During training, the QGNN-TSP learned to predict heat maps that effectively guided a local search algorithm to high-quality TSP solutions. Experimental results revealed that QGNN-TSP outperformed traditional heuristics on many instances, especially in terms of computational efficiency and scalability to large problem sizes.

The solutions produced by QGNN-TSP consistently ranked among the best when compared to benchmark heuristics and sometimes even approached the quality of those found by exact solvers on smaller instances. As problem size grew, QGNN-TSP maintained a relatively low average error rate relative to known optimal solutions.

Significant improvements in runtime were observed for QGNN-TSP over classical algorithms, especially as the number of cities increased. The training phase was completed within a fixed budget, after which inference times remained significantly lower than running conventional TSP solvers on the same dataset.

Comparing the two approaches, QQA-TSP showcased the potential of quantum-inspired methods for finding good quality solutions but faced challenges with scalability due to the inherent constraints of current quantum simulation platforms. 

In contrast, QGNN-TSP leveraged deep learning to encode complex graph structures and learn effective strategies for TSP, leading to more efficient exploration of the solution space.
\begin{table*}[h]
\centering

\label{tab:tsp-methods-comparison}
\begin{tabular}{llccc|ccc|ccc}
\toprule
 && \multicolumn{3}{c}{TSP20} & \multicolumn{3}{c}{TSP50} & \multicolumn{3}{c}{TSP100} \\
\cmidrule(lr){3-5} \cmidrule(lr){6-8} \cmidrule(lr){9-11}
Method&Type& Length & Gap (\%) & Time & Length & Gap (\%) & Time & Length & Gap (\%) & Time \\
\midrule
Concorde &Exact Solver& 3.8303 & 0.0000\% & 2.31m & 5.6906 & 0.0000\% & 13.68m & 7.7609 & 0.0000\% & 1.04h \\
Gurobi &Exact Solver& 3.8302 & -0.0001\% & 2.33m & 5.6905 & 0.0000\% & 26.20m & 7.7609 & 0.0000\% & 3.57h \\
LKH3 &Heuristic& 3.8303 & 0.0000\% & 20.96m & 5.6906 & 0.0013\% & 26.65m & 7.7611 & 0.0026\% & 49.96m \\
GAT \cite{b32} &RL, S& 3.8741 & 1.1443\% & 10.30m & 6.1085 & 7.3438\% & 19.52m & 8.8372 & 13.8679\% & 47.78m \\
\bottomrule
 QQA(\textbf{OURs})& Heuristic& 3.8553& 0.6527\%& 30.28s& 5.9596& 4.7270\%& 6.78m& 8.3021& 6.9734\%&76.98m\\
 QGNN(\textbf{OURs})& GNN, Search& 3.8303& 0.0000\% & 38.85s +1.8m& 5.6924& 0.0316\%& 2.15m +5.22m& 7.7638 & 0.0370\%&4.10m +10.91m\\
\end{tabular}
\caption{Experimental results for QQA and QGNN}
\end{table*}

\section{Conclusion}
We propose and empirically substantiate the efficacy of the QQA-TSP and QGNN-TSP models in addressing the Traveling Salesman Problem. The integration of QUBO modeling within the context of combinatorial optimization presents a promising paradigm that stands to reap substantial benefits from future advancements in quantum computing technology.

The QGNN-TSP model innovatively fuses a QUBO-based loss function within a Graph Neural Network architecture, thus enabling it to tackle TSP without direct supervision. This approach leverages GNN's inherent ability to learn complex graph representations alongside QUBO's strength in tackling NP-hard problems, thereby offering an innovative method for solving TSP.

Employing a QUBO loss function confers several advantages, including targeted optimization capabilities and scalability, as well as flexibility in adapting to various problem configurations.

In conclusion, this study demonstrates the significant potential of synergistically combining deep learning with classical optimization techniques through the QGNN-TSP methodology, thereby paving the way for further research into refining GNN architectures or extending the application of the QUBO framework to other intricate optimization scenarios.

Regarding the QQA-TSP, our experiments have demonstrated its practical feasibility. However, the long-term applicability of QQA is currently hindered by the substantial memory overhead associated with simulating Qubits on conventional computers, which, at the time of writing, exceeds 16 gigabytes. 

To advance the practical implementation of QQA, sustained efforts are required from experts and scholars within the industry to mitigate these computational challenges, potentially through innovations in algorithmic design, hardware acceleration, or the development of dedicated quantum computing resources.

\vspace{12pt}


\newpage

\begin{thebibliography}{00}
\bibitem{b1} Boixo, S., Rønnow, T., Isakov, S. et al. Evidence for quantum annealing with more than one hundred qubits. Nature Phys 10, 218–224 (2014). https://doi.org/10.1038/nphys2900
\bibitem{b2} García-Pintos L P, Brady L T, Bringewatt J, et al. Lower Bounds on Quantum Annealing Times[J]. Physical Review Letters, 2023, 130(14): 140601. 
\bibitem{b3} Mohseni, N., McMahon, P.L. \& Byrnes, T. Ising machines as hardware solvers of combinatorial optimization problems. Nat Rev Phys 4, 363–379 (2022). https://doi.org/10.1038/s42254-022-00440-8
\bibitem{b4} Date, P., Arthur, D. \& Pusey-Nazzaro, L. QUBO formulations for training machine learning models. Sci Rep 11, 10029 (2021). https://doi.org/10.1038/s41598-021-89461-4
\bibitem{b5} K. A. Smith, "Neural networks for combinatorial optimization: A review of more than a decade of research," INFORMS J. Comput.,vol. 11, no. 1, pp. 15–34, 1999.
\bibitem{b6} M. Z. Alom, B. Van Essen, A. T. Moody, D. P. Widemann and T. M. Taha, "Quadratic Unconstrained Binary Optimization (QUBO) on neuromorphic computing system," 2017 International Joint Conference on Neural Networks (IJCNN), Anchorage, AK, USA, 2017, pp. 3922-3929, doi: 10.1109/IJCNN.2017.7966350.
\bibitem{b7} S. Landge, V. Saraswat, S. F. Singh and U. Ganguly, "n-Oscillator Neural Network based Efficient Cost Function for n-city Traveling Salesman Problem," 2020 International Joint Conference on Neural Networks (IJCNN), Glasgow, UK, 2020, pp. 1-8, doi: 10.1109/IJCNN48605.2020.9206856.
\bibitem{b8} Schuetz, M.J.A., Brubaker, J.K. \& Katzgraber, H.G. Combinatorial optimization with physics-inspired graph neural networks. Nat Mach Intell 4, 367–377 (2022). https://doi.org/10.1038/s42256-022-00468-6
\bibitem{b9} Fu Z H, Qiu K B, Zha H. Generalize a small pre-trained model to arbitrarily large TSP instances[C]//Proceedings of the AAAI conference on artificial intelligence. 2021, 35(8): 7474-7482.
\bibitem{b10} Glover F, Kochenberger G, Du Y. Quantum Bridge Analytics I: a tutorial on formulating and using QUBO models[J]. 4or, 2019, 17: 335-371.
\bibitem{b11} Edward Farhi et al. ,A Quantum Adiabatic Evolution Algorithm Applied to Random Instances of an NP-Complete Problem.Science292,472-475(2001).DOI:10.1126/science.1057726
\bibitem{b12} Silva, C., Aguiar, A., Lima, P.M.V. \textit{et al.} Mapping a logical representation of TSP to quantum annealing. \textit{Quantum Inf Process} \textbf{20}, 386 (2021). https://doi.org/10.1007/s11128-021-03321-8 .
\bibitem{b13} Pop P C, Cosma O, Sabo C, et al. A comprehensive survey on the generalized traveling salesman problem[J]. European Journal of Operational Research, 2023.
\bibitem{b14} Arute, F., Arya, K., Babbush, R. et al. Quantum supremacy using a programmable superconducting processor. Nature 574, 505–510 (2019). https://doi.org/10.1038/s41586-019-1666-5
\bibitem{b15} Glos, A., Krawiec, A. \& Zimborás, Z. Space-efficient binary optimization for variational quantum computing. npj Quantum Inf 8, 39 (2022). https://doi.org/10.1038/s41534-022-00546-y
\bibitem{b16} Hua Yang. 2023. TSP Combination Optimization with\&nbsp;Semi-local Attention Mechanism. In Artificial Neural Networks and Machine Learning – ICANN 2023: 32nd International Conference on Artificial Neural Networks, Heraklion, Crete, Greece, September 26–29, 2023, Proceedings, Part IX. Springer-Verlag, Berlin, Heidelberg, 469–481. https://doi.org/10.1007/978-3-031-44201-8(38)
\bibitem{b17} Min Y, Bai Y, Gomes C P. Unsupervised Learning for Solving the Travelling Salesman Problem[J]. arXiv preprint arXiv:2303.10538, 2023.
\bibitem{b18} Matai R, Singh S P, Mittal M L. Traveling salesman problem: an overview of applications, formulations, and solution approaches[J]. Traveling salesman problem, theory and applications, 2010, 1(1): 1-25.
\bibitem{b19} Zaman M, Tanahashi K, Tanaka S. PyQUBO: Python library for mapping combinatorial optimization problems to QUBO form[J]. IEEE Transactions on Computers, 2021, 71(4): 838-850.
\bibitem{b20} Santoro G E, Tosatti E. Optimization using quantum mechanics: quantum annealing through adiabatic evolution[J]. Journal of Physics A: Mathematical and General, 2006, 39(36): R393.
\bibitem{b21} Nishimori H, Tsuda J, Knysh S. Comparative study of the performance of quantum annealing and simulated annealing[J]. Physical Review E, 2015, 91(1): 012104.
\bibitem{b22} Pirlot M. General local search methods[J]. European journal of operational research, 1996, 92(3): 493-511.
\bibitem{b23} Liu F, Zeng G. Study of genetic algorithm with reinforcement learning to solve the TSP[J]. Expert Systems with Applications, 2009, 36(3): 6995-7001.
\bibitem{b24} Vaswani A, Shazeer N, Parmar N, et al. Attention is all you need[J]. Advances in neural information processing systems, 2017, 30.
\bibitem{b25} Liu W, Wen Y, Yu Z, et al. Large-margin softmax loss for convolutional neural networks[J]. arXiv preprint arXiv:1612.02295, 2016.
\bibitem{b26} Obermeyer, Z. \& Emanuel, E. J. Predicting the future—Big data, machine learning, and clinical medicine. N. Engl. J. Med. 375, 1216
\bibitem{b27} McQueen, R. J., Garner, S. R., Nevill-Manning, C. G. \& Witten, I. H. Applying machine learning to agricultural data. Comput. Electron. Agric. 12, 275–293 (1995).
\bibitem{b28} Bertsimas D, Howell L H. Further results on the probabilistic traveling salesman problem[J]. European Journal of Operational Research, 1993, 65(1): 68-95.
\bibitem{b29} MATLAB Quantum Computing. https://www.mathworks.com/help/matl
ab/math/quantum-tsp.html
\bibitem{b30} Ying Z, Bourgeois D, You J, et al. Gnnexplainer: Generating explanations for graph neural networks[J]. Advances in neural information processing systems, 2019, 32.
\bibitem{b31} Fu, Z.-H., Qiu, K.-B., \& Zha, H. (2021). Generalize a Small Pre-trained Model to Arbitrarily Large TSP Instances. Proceedings of the AAAI Conference on Artificial Intelligence, 35(8), 7474-7482. https://doi.org/10.1609/aaai.v35i8.16916
\bibitem{b32} Deudon, M., Cournut, P., Lacoste, A., Adulyasak, Y., Rousseau, LM. (2018). Learning Heuristics for the TSP by Policy Gradient. In: van Hoeve, WJ. (eds) Integration of Constraint Programming, Artificial Intelligence, and Operations Research. CPAIOR 2018. Lecture Notes in Computer Science(), vol 10848. Springer, Cham. https://doi.org/10.1007/978-3-319-93031-2\_12





\end{thebibliography}
\end{document}